# Porous Materials with Omnidirectional Negative Poisson's Ratio

*Giorgio Carta, Michele Brun\*, and Antonio Baldi*

Department of Mechanical, Chemical and Materials Engineering, University of Cagliari,

Piazza d'Armi, 09123 Cagliari, Italy

E-mail: mbrun@unica.it

**Abstract**

This paper presents an auxetic medium, consisting of a two-dimensional perforated sheet where the holes are arranged in a repetitive pattern. The hexagonal disposition of the perforations makes the medium isotropic in the plane. It is shown that negative values of the Poisson's ratio can be achieved for specific values of the dimensions and orientations of the holes. The results of the numerical simulations are confirmed by experimental tests, in which the Poisson's ratio of each specimen is evaluated from the displacement field obtained from the Digital Image Correlation (DIC) technique. The distribution of stresses in the medium is determined directly from photoelastic images. The auxetic structure proposed in this paper is easy to fabricate and can be very useful in several engineering applications.

Keywords: auxetic, negative Poisson's ratio, metamaterials, isotropy, digital image correlation

## 1. Introduction

Most natural materials are characterized by a positive Poisson's ratio, namely they are observed to contract (expand) laterally when stretched (compressed) longitudinally. Nonetheless, the classical theory of elasticity does not preclude the existence of materials with negative Poisson's ratio, known also as "auxetics" after Evans.[1] In particular, for isotropic materials thermodynamic stability arguments lead to the condition that the Poisson's ratio lies in the interval (-1, 0.5),[2] while for anisotropic materials this interval is unbounded.

The first example in the literature of an auxetic material was provided by Love,[3] who calculated the Poisson's ratio of cubic crystals of pyrite as nearly equal to -1/7. Natural materials exhibiting a negative Poisson's ratio include silicates,[4] cubic elemental metals,[5] zeolites[6,7] and ceramics.[8] Lakes[9] presented the first designed auxetic material, consisting of a reentrant foam. Since this trailblazing work many artificially-made auxetic models have been proposed, based on different mechanisms such as reentrant units,[10,11] chirality,[12-14] rigid rotating units[15,16] and elastic instability.[17,18] Recently, Taylor et al.[19] showed that metallic sheets with orthogonal elliptical voids, having two-dimensional cubic symmetry, are auxetic for specific values of the porosity and of the voids' shape. Chen et al.[20] carried out a numerical and experimental investigation on a fiber-reinforced composite flexible skin with in-plane negative Poisson's ratio. Chen et al.[21] studied a curved cellular structure, manufactured by using Kirigami techniques, that is characterized by a null Poisson's ratio. Cabras and Brun[22] analyzed and fabricated two-dimensional lattice structures, with isotropic or cubic behaviors, having a Poisson's ratio close to -1. Other models are described in recent reviews.[23-27]

In comparison with traditional materials, auxetic systems have some superior characteristics that can be exploited technologically, like higher resistance to indentation, larger fracture toughness and enhanced vibration absorption properties.[16] Relevant applications where auxetic systems may be particularly useful are the replacement of blood vessels,[28] the fabrication of smart filters and fasteners,[9] the design of structures with double curvature for aircraft wings and car doors[28,29] and the production of auxetic fibers that are more resistant to pull-out.[28]

Here we present a porous auxetic structure, which can be made of any material. We determine the Poisson's ratio by performing experimental tests on different specimens and we compare the results with numerical simulations. In addition, we investigate how the value of the Poisson's ratio changes with the orientation angle and the dimensions of the holes. We envisage that this structured medium can be of great value to devise novel stents and artificial blood vessels in biomedical engineering[28] and to design the structural components of gas turbines constituted of perforated surfaces.[19]

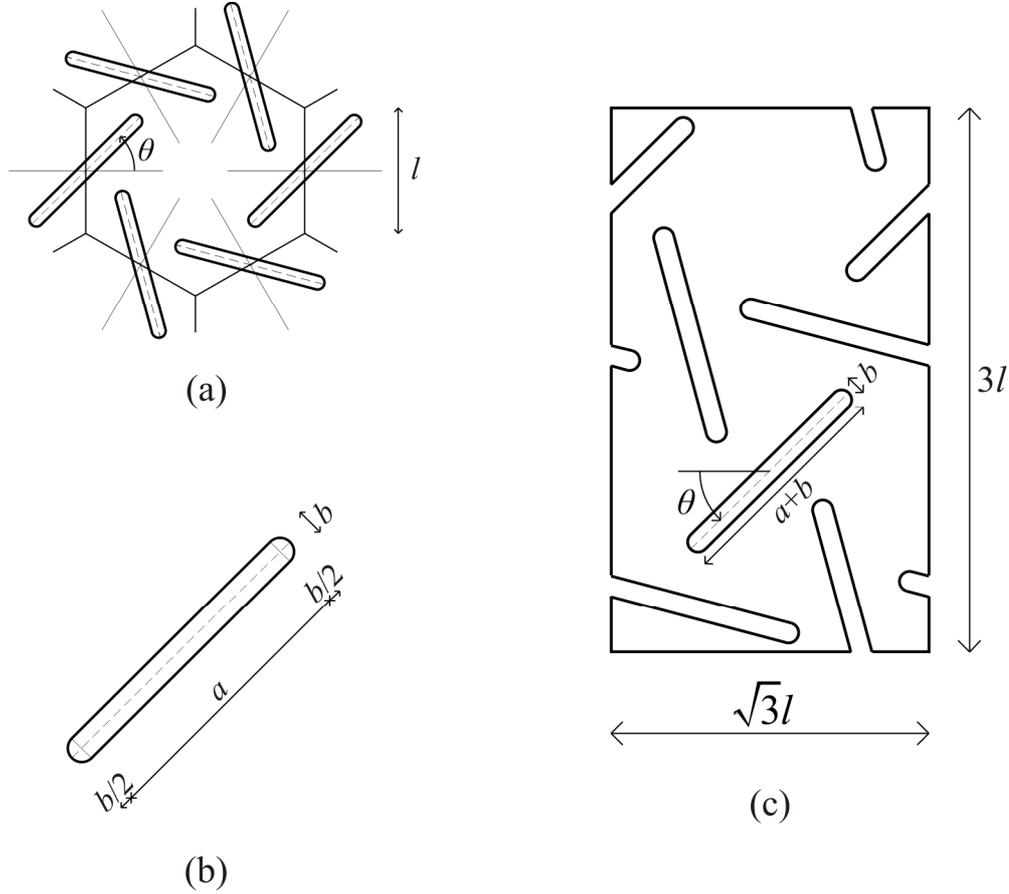

**Figure 1.** (a) Disposition of the perforations inside the two-dimensional structure, where $l$ is the side of the hexagon and $\theta$ is the orientation angle of the perforations with respect to the normals to the hexagonal sides; (b) detail of each perforation, made of an elongated rectangle ending with two semicircles; (c) elementary cell representative of the infinite periodic structure.

## 2. Results and discussion

### 2.1. Description of the model

We study a two-dimensional sheet containing holes, arranged in a hexagonal pattern as shown in Figure 1a. The holes are inclined by an angle $\theta$ with respect to the normals to the hexagon sides, which have length $l$. As sketched in Figure 1b, each hole consists of a rectangle with two semicircles at the ends in order to avoid stress concentration.

The novelty of this model with respect to other porous structures, for example the one analyzed by Taylor et al.,[19] lies in its planar isotropic behavior, ensuing from the 60°-symmetry of the voids' arrangement. The system described by Taylor et al.[19] is instead anisotropic, since it exhibits a cubic symmetry. Furthermore, we present a thorough study of the elastic modulus of the homogenized structure, which is highly relevant for practical applications.

## 2.2. Experimental findings

We test three specimens, indicated with A, B and C in Figure 2 and in the rest of the paper. They are made of Lexan polycarbonate and have a length of 420 mm, a width of 110 mm and a thickness of 3 mm. The geometry of the perforations is identified by the following quantities: $a = 10$ mm, $b = 1$ mm, $l \approx 9$ mm and $\theta = 45°$. Samples B and C differ from sample A since the voids' arrangement in B and C is rotated by 45° and 90°, respectively, relatively to A. This allows to test the system along three different directions if, for all samples, the load is applied along the length of the specimen, namely in the $y$ direction.

The constitutive properties of the intact material are determined from a simple tensile test on homogeneous samples with the supplementary use of a strain gauge to measure lateral displacements. The Young's modulus, Poisson's ratio and yield stress are found to be equal to *E = 2.9 GPa, v = 0.35* and *σy = 40 MPa*, respectively.

We obtain the displacement and strain fields within the samples by means of the Digital Image Correlation (DIC) technique. Digital Image Correlation is a non-interferometric experimental technique able to estimate bi-dimensional (tri-dimensional in its stereo version) displacement fields. Its basic assumption is the constancy of the intensity of each point during motion, thus, if the surface texture is not uniform, it is possible to correlate each location of the reference image with a corresponding displaced point in the test one. Formally, this results in the well-known optical flow equation

$$\frac{\partial I}{\partial x}\dot{u} + \frac{dI}{dy}\dot{v} + \frac{dI}{dt} = 0,$$

where u and v are the x and y displacement components and I is the image intensity. Looking to this equation it is apparent that it involves two unknowns, thus an auxiliary condition has to be specified. The most commonly used solution to this problem is the Lucas-Kanade approach[30], which assumes the displacement field can be locally approximated using a simple function (usually an affine transform). Providing a sufficiently large number of pixels is involved in the least square fit, the set of parameters controlling the mapping can be easily identified by reverse calibration (usually by a Newton-like iterative algorithm); moreover, by sampling the images at different locations, full-field data can be obtained by interpolating the displacements at the sampling points.

From the practical viewpoint, performing the measurement requires acquisition of two images before and after load applications. To comply with the basic hypothesis of the technique[31] (pixel intensity does not change with motion) illumination has to be (nearly) isotropic, specimen surface has to be opaque (to minimize directional components associated with reflection) and the surface has to be textured with a chaotic pattern (to ensure its uniqueness).

For each sample, we consider a rectangular area in the central part of the specimen, where the boundary effects are negligible. We calculate the averages of the horizontal displacements (*u*) in the left (L) and right (R) sides of the rectangle and the averages of the vertical displacements (*v*) in the bottom (B) and top (T) boundaries.

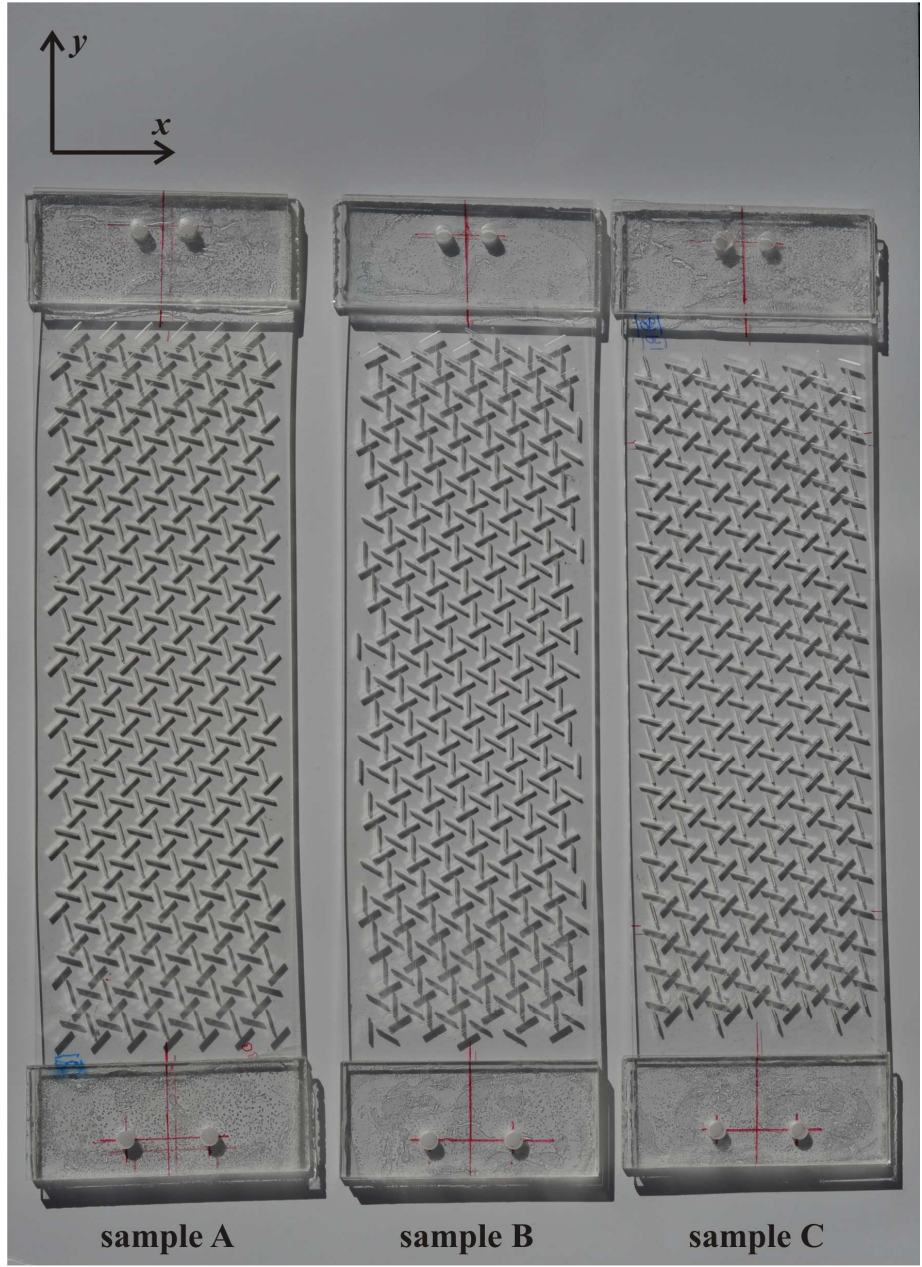

**Figure 2.** Specimens used in the experimental tests, where $a = 10$ mm, $b = 1$ mm, $l \approx 9$ mm and $\theta = 45°$ (refer to Figure 1 for the definitions of the geometrical quantities). The disposition of the holes in samples B and C is rotated by 45° and 90°, respectively, with respect to the disposition in sample A.

Accordingly, the experimental value of the Poisson's ratio is estimated as

$$\nu^{\exp} \approx -\frac{<\varepsilon_{xx}>}{<\varepsilon_{yy}>} = -\frac{\left(<u^{R}> - <u^{L}>\right)/L_x}{\left(<v^{T}> - <v^{B}>\right)/L_y}, \qquad (1)$$

where $<q>$ represents the average value of the quantity $q$, $L_x$ and $L_y$ are the dimensions of the horizontal and vertical sides of the rectangle, while $\varepsilon_{xx}$ and $\varepsilon_{yy}$ stand for the axial strains in the *x* and *y* directions. We also consider different values of the applied load, yielding different strains $\varepsilon_{yy}$, but restricting the range of deformation within the linear regime.

We report the results for the three specimens and for three values of the applied load in Figure 3. The experimental values of $v^{\text{exp}}$ are indicated by black dots. We note that, as $\varepsilon_{yy}$ increases, the data tend to converge to the horizontal dashed line, which indicates the theoretical value determined from the periodic analysis described in Section 2.4. Thus, the load needs to be sufficiently large to allow for the necessary adjustment of the specimen. Figure 3 shows that the proposed structured medium is auxetic, since the Poisson's ratio is negative. In addition, the latter does not depend on the loading direction, hence the material is isotropic in the plane. By passing, we note that the experiment along the three different directions are sufficient in order to verify the isotropy of the heterogeneous medium.

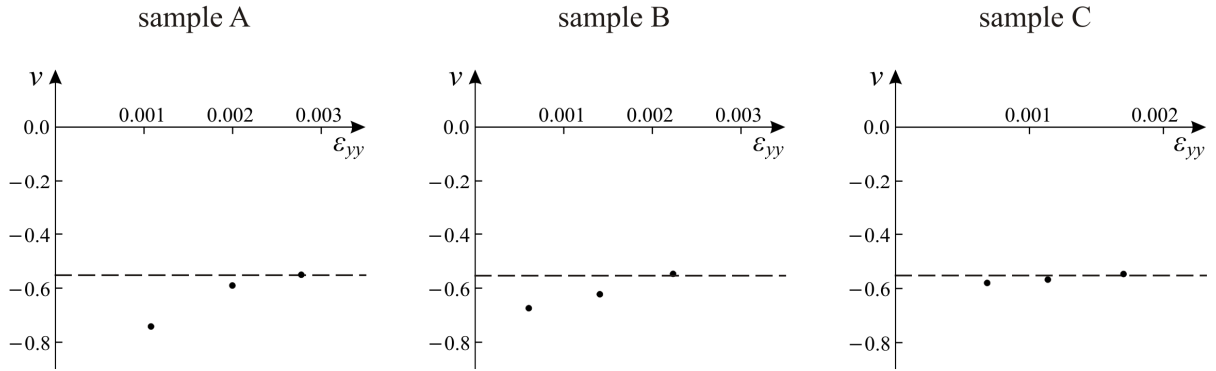

**Figure 3.** Experimental values of the Poisson's ratio for the samples A, B and C in Figure 2, obtained for different values of the axial strain $\varepsilon_{yy}$. The horizontal dashed line $v = -0.553$ represents the theoretical value of the Poisson's coefficient for the infinite periodic structure.

**2.3. Numerical computations for the finite structure**

We simulate the tests numerically by means of a two-dimensional finite element model developed in *Comsol Multiphysics* (version 4.3b). The mesh consists of 95690 triangular elements and is refined in proximity of the holes. We apply a uniform vertical load at the top boundary and impose zero vertical displacements at the bottom boundary and zero horizontal displacements at the middle nodes of both horizontal boundaries in order to prevent rigid body motion. We consider a rectangular area in the central part of the model, where we compute the average values of the normal stresses $\sigma_{xx}$ and $\sigma_{yy}$ and of the axial strains $\varepsilon_{xx}$ and $\varepsilon_{yy}$. We solve the problem by performing a plane stress analysis, which is justified by the loading conditions and the small thickness. Simple calculations lead to the following expressions for the Poisson's coefficient $v$ and the Young's modulus $E$:

$$v = v_{xy} = \frac{\varepsilon_{xx}\sigma_{yy} - \varepsilon_{yy}\sigma_{xx}}{\varepsilon_{xx}\sigma_{xx} - \varepsilon_{yy}\sigma_{yy}} ; \tag{2}$$

$$E = E_x = E_y = \frac{\sigma_{xx}^2 - \sigma_{yy}^2}{\varepsilon_{xx}\sigma_{xx} - \varepsilon_{yy}\sigma_{yy}} . \tag{3}$$

We report the results in Table 1, together with the outcomes obtained from the experiments and from the periodic study discussed in Section 2.4.

**Table 1.** Values of Poisson's ratio $v$ and Young's modulus $E$ determined from the experiments and from the numerical computations on the finite and periodic structures.

|  | experimental | numerical (finite) | numerical (periodic) |  |
|---|---|---|---|---|
| sample A | -0.551 | -0.552 | -0.553 | $v$ |
|  | --- | 248.8 | 254.8 | $E$ (MPa) |
| sample B | -0.550 | -0.552 | -0.553 | $v$ |
|  | --- | 247.8 | 254.8 | $E$ (MPa) |
| sample C | -0.548 | -0.557 | -0.553 | $v$ |
|  | --- | 247.0 | 254.8 | $E$ (MPa) |

The comparison of the Poisson's ratios of the three samples reveals an excellent agreement between the numerical and experimental data. We observe that the numerical findings corresponding to the three finite models are very close to each other, hence the material behavior is isotropic in the plane; clearly, the small discrepancies are due to the tiny - but not vanishing - boundary effects. We have also checked that by using Equation 1 and the relative procedure we obtain values of the Poisson's ratio close to those indicated in the second column of Table 1.

In Figure 4 we plot the contour maps of the horizontal (a) and vertical (b) displacements of the sample A, obtained both experimentally and numerically. It is apparent that the correspondence between the two studies is excellent.

The internal black regions represent the perforations, for which the displacement field could not be evaluated. The same comparative analysis for the samples B and C is reported in the supplementary material.

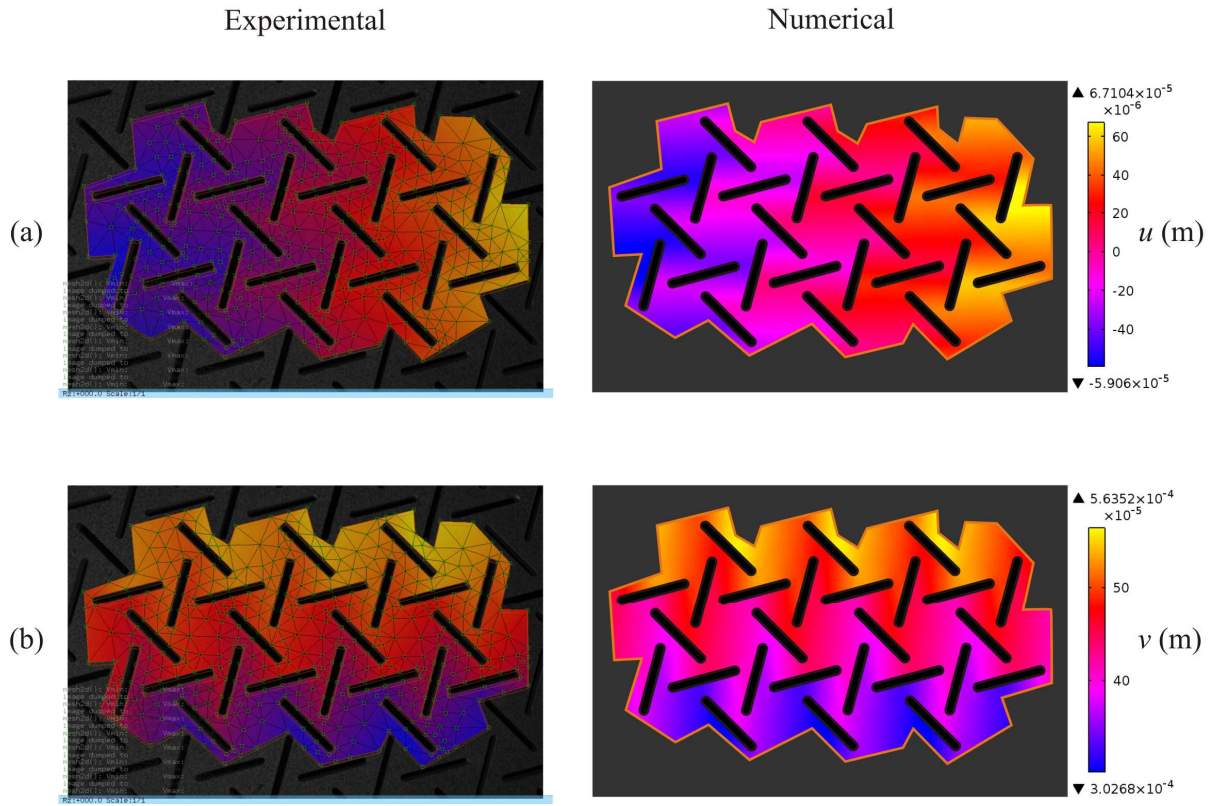

**Figure 4.** Contour maps of the horizontal (a) and vertical (b) components of the displacement field, obtained from the lab tests and from the finite element simulations, for the sample A displayed in Figure 2. The average axial strain is $\varepsilon_{yy} = 2.77 \cdot 10^{-3}$.

**2.4. Periodic analysis**

We assume that the perforated sheet is of infinite extent and made of repetitive cells, as that shown in Figure 1c. We construct a finite element model of the unit cell in *Comsol Multiphysics* by using a discretization of 10921 triangular elements. We impose periodic conditions at the boundaries of the cell and we fix three degrees of freedom to avoid rigid body motion.

We apply firstly an axial strain $\varepsilon_{xx} = 10^{-4}$, secondly an axial strain $\varepsilon_{yy} = 10^{-4}$ and finally a shear strain $\varepsilon_{xy} = 10^{-4}$. For all three cases, we compute the average values of the relevant components of the stress and strain tensors. We determine the Poisson's coefficient and the Young's modulus from Equation 2 and Equation 3, respectively, while the shear modulus is given by

$$\mu = \mu_{xy} = \frac{\sigma_{xy}}{2\varepsilon_{xy}}. \tag{4}$$

The first and second cases ($\varepsilon_{xx} = 10^{-4}$ and $\varepsilon_{yy} = 10^{-4}$) yield the same results: $\nu = -0.553$ and $E = 2.548 \cdot 10^8$ Pa. From the third case ($\varepsilon_{xy} = 10^{-4}$) we obtain $\mu = 2.849 \cdot 10^8$ Pa, which is equal to $E/[2(1+\nu)]$. This analysis confirms that the medium is auxetic and isotropic.

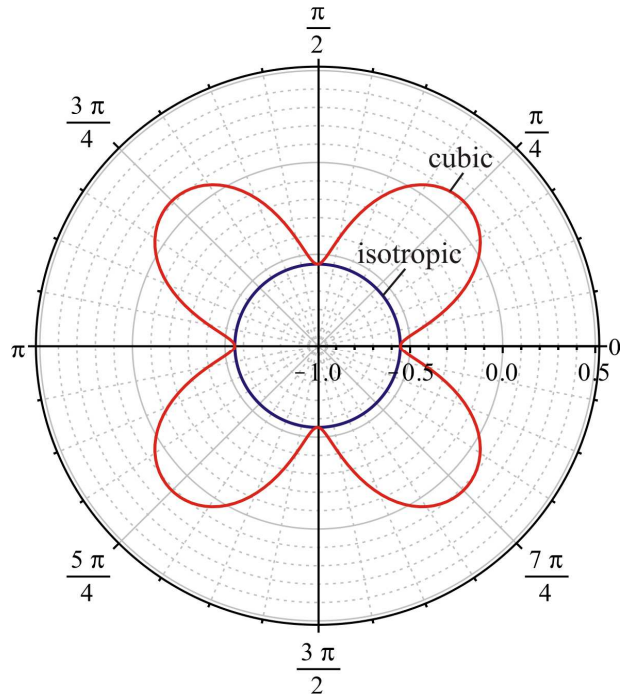

**Figure 5.** Polar diagrams of the Poisson's ratio as a function of the angle $\beta$, defining the orientation of the stretching direction with respect to a principal axis of the structure, for the model described in this paper (blue color) and for a generic cubic material with $\mu \neq E/[2(1+\nu)]$ (red color).

For a generic cubic material, the Poisson's ratio depends on the stretching direction. Denoting by $\beta$ the angle between the stretching direction and a principal axis of the structure, the Poisson's ratio can be calculated by using the following formula:[22]

$$\nu(\beta) = -\frac{\cos^2(\beta)\sin^2(\beta)(2/E - 1/\mu) - \left[\sin^4(\beta) + \cos^4(\beta)\right]\nu/E}{\left[\sin^4(\beta) + \cos^4(\beta)\right]/E + 1/\mu - 2\nu/E}, \quad (5)$$

where $E$ and $\nu$ are determined from a tensile test and $\mu$ from a shear test.

In Figure 5 we show the polar diagrams of $\nu(\beta)$ for the structured medium proposed in this paper and for a generic cubic material. It is clear that the present model is isotropic in the plane, since $\nu$ does not vary with $\beta$. On the other hand, the Poisson's ratio of a generic cubic material depends on $\beta$, and it can even be negative in some directions but positive in others. Obviously, the variation of $\nu$ with the direction of the applied load can have significant drawbacks in practical applications.

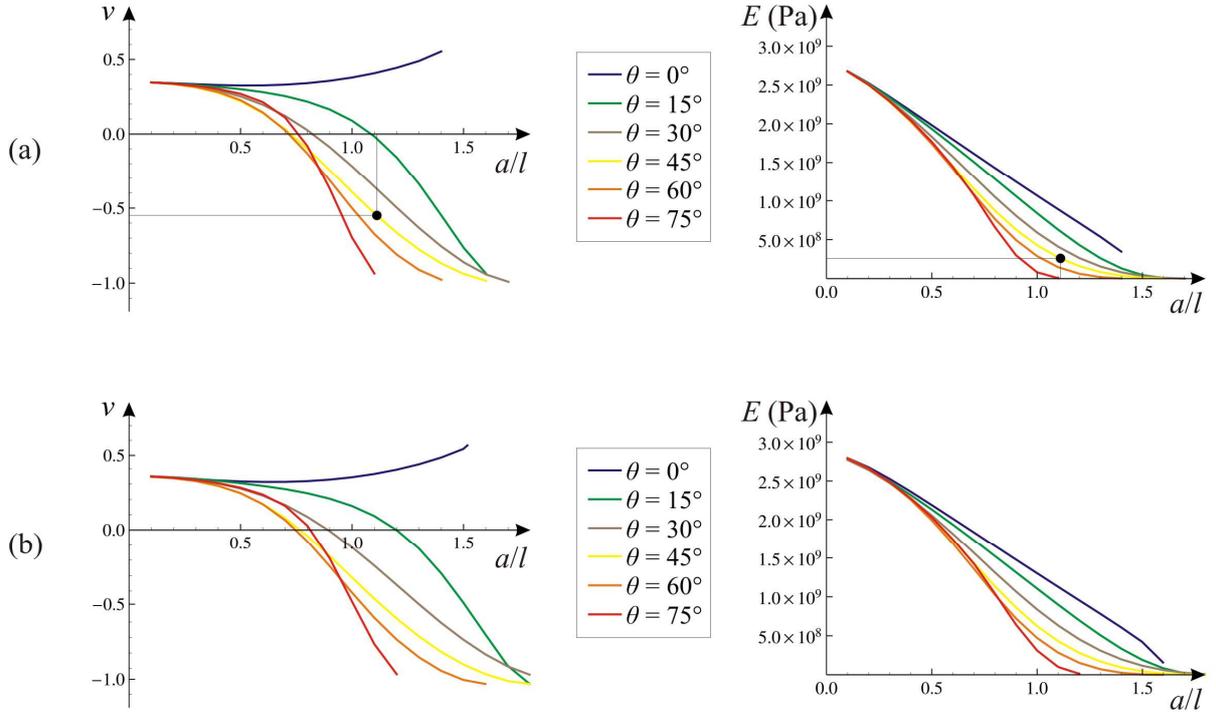

**Figure 6.** Poisson's ratio $\nu$ and elastic modulus $E$ as functions of $a/l$ for different values of $\theta$ and for $b/l = 1/9$ (a) and $b/l = 1/18$ (b). The dots in the inset (a) indicate the values of $\nu$ and $E$ for the samples in Figure 2 (see also Table 1).

We perform a parametric study to investigate how the ratios $a/l$ and $b/l$ and the orientation angle of the voids $\theta$ affect the constitutive properties of the perforated sheet. In Figure 6 we show how the Poisson's ratio and the Young's modulus vary with the ratio $a/l$ for different values of $\theta$ and for $b/l = 1/9$ in part (a) and for $b/l = 1/18$ in part (b). The results obtained for the particular values of the parameters characterizing the specimens in Figure 2 are highlighted by black dots in Figure 6a.

We point out that, in order to avoid the merging of the holes, their length *a* must be kept below a certain limit, which depends on the angle $\theta$. This limit is obtained from geometrical considerations and is equal to

$$a_1 = \frac{l\left[\sqrt{3} + 2\sin(2\theta)\right]}{\cos(\theta) + \sqrt{3}\sin(\theta)} - \frac{4\sqrt{3}\,b}{3} \quad \text{for } \theta < \theta^*, \tag{6a}$$

$$a_2 = l\sin(\theta) - \sqrt{\frac{4b^2}{3} - l^2\cos^2(\theta)} \quad \text{for } \theta > \theta^*. \tag{6b}$$

The transition value $\theta^*$ is calculated by solving the equation $a_1(\theta) = a_2(\theta)$. We also note that Equation (6b) requires that $|\cos(\theta)| < 2b/(\sqrt{3}\,l)$.

Figure 6 shows that for low values of *a/l*, namely for short voids, the Poisson's coefficient of the perforated sheet is close to the value of the intact material, equal to 0.35 in this case. As the length of the holes *a* is increased, the value of *v* diminishes and, eventually, the structured medium becomes auxetic ($v < 0$). At the same time, the elastic modulus obviously decreases, hence the material becomes softer. It is interesting - and not trivial - that, generally, *v* and *E* can be decreased by increasing the orientation angle of the voids $\theta$, for a fixed value of *a/l*. The comparison between the insets (a) and (b) reveals that, for specified values of *a/l* and $\theta$, *v* and *E* can be decreased by making the perforations wider.

## 2.5. Photoelasticity

We use photoelasticity to evaluate the stress distribution in the porous medium. In Figure 7a we show the difference between the principal stresses, namely $\sigma_1$-$\sigma_2$, obtained experimentally with a circular polariscope. It is apparent that the structure exhibits high concentration of stress in proximity of the ends of the perforations. Figure 7b presents the $\sigma_1$-$\sigma_2$ distribution derived from the finite element model. Once again, the comparison between the experimental and numerical outcomes is excellent.

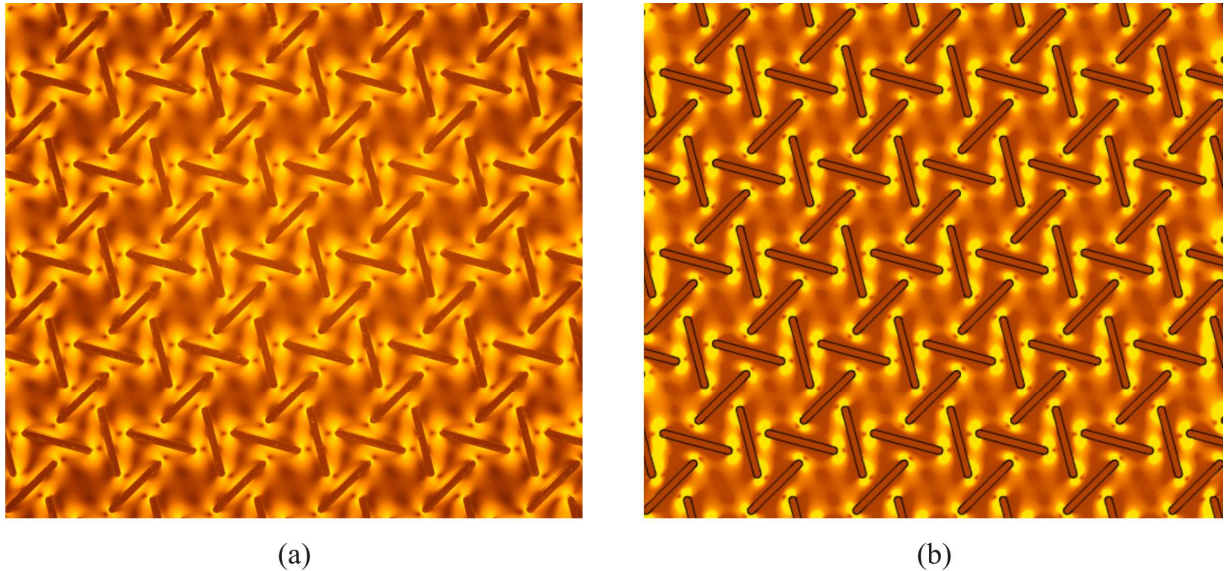

(a)                                                               (b)

**Figure 7.** Contour maps of the difference in the first and second principal stresses ($\sigma_1 - \sigma_2$) obtained experimentally with a circular polariscope (a) and numerically by means of finite element computations (b). The scale goes from yellow (highest value) to dark orange (lowest value).

## 3. Conclusions

In summary, we have shown that a two-dimensional sheet with elongated perforations arranged in a periodic hexagonal pattern exhibits a negative Poisson's ratio, which does not change with the loading direction. The planar auxetic and isotropic behavior of the considered porous medium has been observed experimentally on three specimens, characterized by a 45° rotation of the holes' disposition relative to each other and loaded in the same direction. The experimental findings have been corroborated by numerical computations on finite models, reproducing the real specimens. Very similar results have been obtained from finite element simulations based on the periodic elementary cell of the structure. The latter has also been employed to perform a parametric study of the geometric quantities of the system, which has shown in particular how the Poisson's ratio and the Young's modulus vary with the dimensions and the orientation angles of the perforations. In addition, photoelasticity has been used to determine the stress distribution in the medium. We believe that this structure, which is simple to manufacture and easily adaptable to specific requirements, may be exploited in a great variety of industrial applications, considering that the results of this paper are scale-independent and are valid also for high deformations in the material.

## 4. Experimental Section

The standard deviation of displacements, when acquired using the standard Digital Image Correlation, is relatively large: indeed due to the statistical approach, it depends on the square root of the number of pixels involved in the local computation. Moreover, each sampling is uncorrelated, thus the strain uncertainty is about 250 μm/m. To reduce this value, we replaced the local measurement with a global approach: instead of using a local model of the displacement field, we opted for a global description using a finite-element-like approach[32]. Displacement inside each element are controlled by nodal parameters, which, being shared with adjacent elements, makes the model global. In this way the standard deviation of displacements is significantly reduced[33] and the solution is more robuts.

From the experimental viewpoint, on the basis of Finite Element simulations, we apply a series of 5 displacement steps (0.5 mm each) along the *y* direction taking care not to exceed the yield stress (this is confirmed by the a-posteriori photoelastic inspection). Images are acquired using a D700 Nikon camera with a 200 mm focal lens (to reduce prospective artifacts). To remove the influence of the Bayer pattern, the resulting (raw) images are binned using a 2×2 mask. Then they are processed using an in-house-developed software (ofTri). Finally the output data (nodal displacements and strains) are post-processed to compute the Poisson ratio.

**Acknowledgements**

G. Carta and M. Brun acknowledge the financial support of Regione Autonoma della Sardegna (LR7 2010, Grant 'M4' CRP-27585). The authors wish to thank Mr. Gianluca Marongiu for the preparation of the specimens, and Prof. Francesco Aymerich and Dr. Agostino Cerioni for the tests on the material properties of the intact structure.

**Supporting Information**

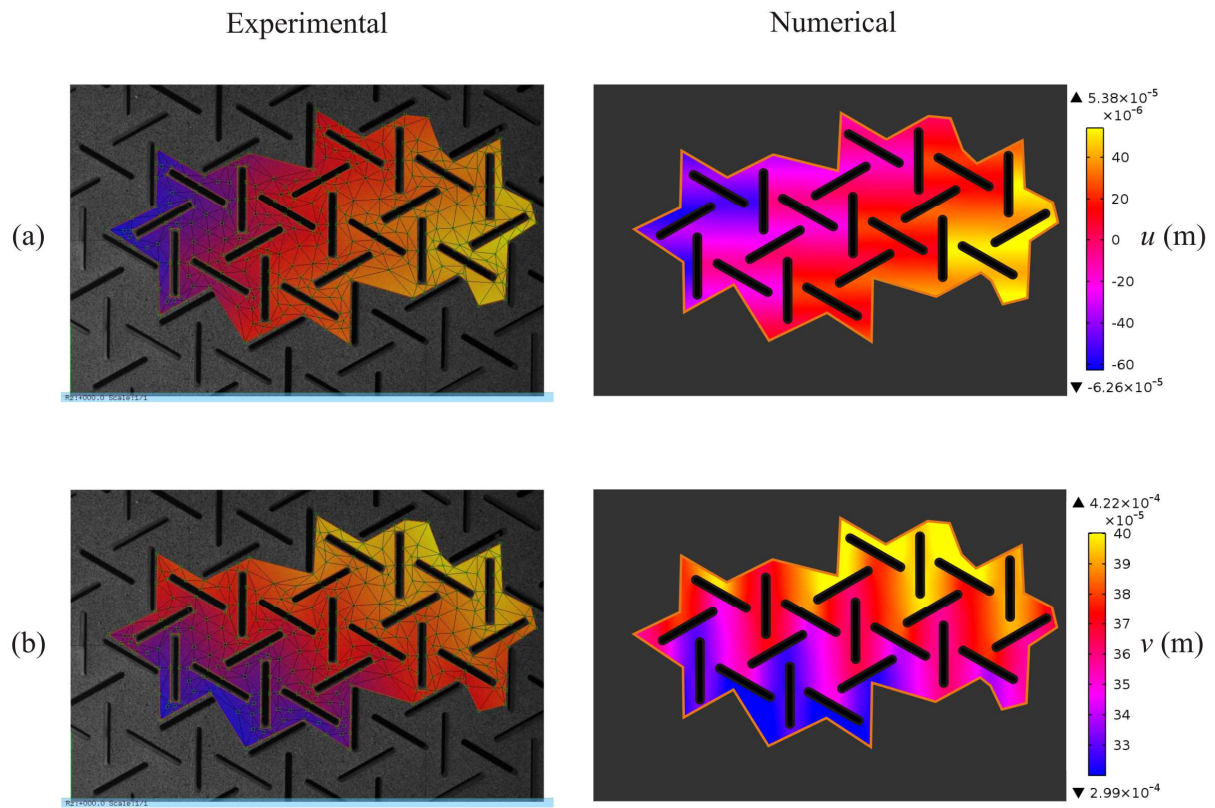

**Figure A.** Contour maps of the horizontal (a) and vertical (b) components of the displacement field, obtained from the lab tests and from the finite element simulations, for the sample B shown in Figure 2. The average axial strain is $\varepsilon_{yy} = 2.24 \cdot 10^{-3}$.

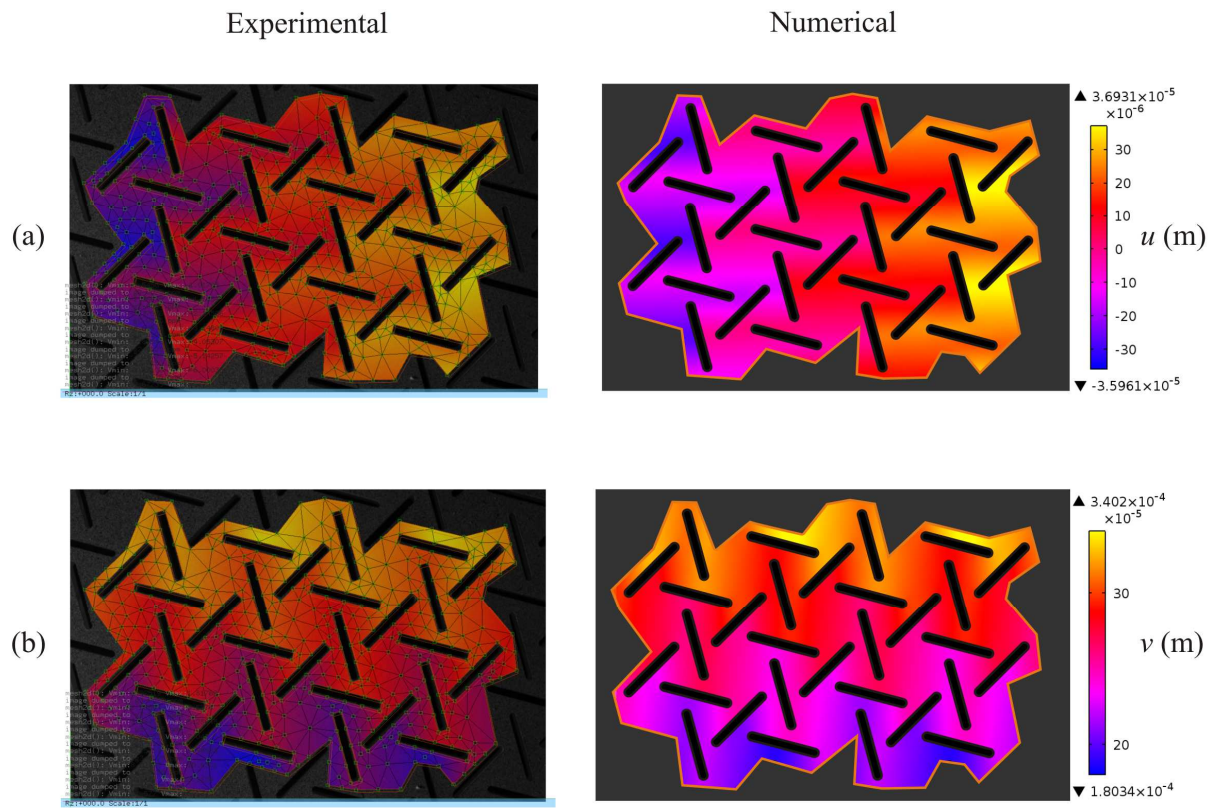

**Figure B.** Contour maps of the horizontal (a) and vertical (b) components of the displacement field, obtained from the lab tests and from the finite element simulations, for the sample C shown in Figure 2. The average axial strain is $\varepsilon_{yy} = 1.70 \cdot 10^{-3}$.